\documentclass[aps,prl,oneside,onecolumn]{revtex4-1}
\usepackage{amsmath}
\usepackage{amssymb}
\usepackage{graphicx}

\begin{document}

\draft \title{Homotopy analysis method applied to second-order frequency mixing in nonlinear optical dielectric media}
\author{Nathan J. Dawson}
\email{ndawson@hpu.edu}
\address{Department of Natural Sciences, Hawaii Pacific University, Kaneohe, HI 96744, USA\\
Department of Computer Science and Engineering, Hawaii Pacific University, Honolulu, HI 96744, USA\\
Department of Physics and Astronomy, Washington State University, Pullman, WA 99164, USA}

\author{Moussa Kounta}
\address{Department of Mathematics, University of the Bahamas, Nassau, Bahamas}

\begin{abstract}The classical problem of three-wave mixing in a nonlinear optical medium is investigated using the homotopy analysis method (HAM). We show that the power series basis builds a generic polynomial expression that can be used to study three-wave mixing for arbitrary input parameters. The phase-mismatched and perfectly phase matched cases are investigated. Parameters that result in generalized sum- and difference-frequency generation are studied using HAM with a power series basis and compared to an explicit finite-difference approximation. The convergence region is extended by increasing the auxiliary parameter.\end{abstract}

\maketitle

\section{Introduction}

Experimental realization of three-wave mixing of coherent light was first reported by Franken et al. in 1961, where second-harmonic generation was observed.\cite{frank61.01} In 1962, Armstrong et al. developed methods to analytically describe second harmonic generation as well as general sum- and difference-frequency generation.\cite{armst62.01} Sum- and difference-frequency generation were observed within the following year using a ruby laser and mercury lamp.\cite{smith62.01,smith63.01}

The field of nonlinear optics typically focuses on either the macroscopic observations such as self focusing \cite{akhma68.01} and nonlinear absorption,\cite{kaise61.01} or the fundamental properties of materials that underpin the nonlinear optical response.\cite{orr71.01} The electronic response often involves numerical techniques to theoretically determine the strength of the nonlinear electronic response of materials in the quantum regime,\cite{agren93.01,karna93.01,jonss06.01} although some analytical tools have been developed to better understand the limits of the nonlinear-optical coefficients to design better materials.\cite{kuzyk13.01,lytel15.02,dawson15.01,dawso16.01,lytel17.01,dawso18.01} The macroscopic phenomena predicted by classical electromagnetism equations contain the nonlinear-optical coefficients determined from their microscopic properties.\cite{bloem96.01,meyst07.01,xiong11.01} Mechanisms other than the electronic response can result in nonlinear optical phenomena such as the vibrational response,\cite{chou09.01} molecular reorientation,\cite{wong74.01} and thermo-optic effect,\cite{kovsh99.01} albeit the electronic response time is quicker.\cite{chris10.01,kuzyk13.02}

The clever and yet complicated description of general three-wave mixing provided by Armstrong et al. required the constraints of power flow equations in addition to the three nonlinear amplitude equations.\cite{armst62.01} The solution to the three-wave mixing problem also involved ranking the roots of a cubic equation. The solutions to the nonlinear amplitudes were based on the Jacobi elliptic $sn$ function and contain the roots of the cubic equation both inside and outside of the special function's argument. The undepleted pump approximation and other special cases have been used to formulate simplified expressions for specific sets of parameters.\cite{boyd09.01} Numerical methods can also be used to quickly approximate the slow-varying field amplitudes such as the explicit finite-difference scheme.

The homotopy analysis method (HAM) was developed to approximate nonlinear differential equations using analytical expressions.\cite{liao12.01} The HAM has previously been used to describe the behavior of a pulse propagating in a semiconducting optical amplifier.\cite{jia18.01} Analytical expressions for time-dependent eikonal equations \cite{dehgh11.01} and the nonlinear Schrodinger equation \cite{dehgh10.01} have also been generated using the HAM. In this paper, we show that the classical problem of three-wave mixing in a second-order material can be approximated in terms of common functions using the HAM.

\section{Review of Wave Propagation in Nonlinear Dielectric Media}
\label{sec:sumfreqgen}

The time-domain wave equation for the electric field of a light wave propagating in a nonlinear dielectric medium is given by
\begin{equation}
\nabla^2 \vec{E}\left(\vec{r},t\right) - \nabla\left(\nabla \cdot \vec{E}\left(\vec{r},t\right)\right) = \frac{1}{c^2}\frac{\partial^2 \vec{E}\left(\vec{r},t\right)}{\partial t^2} + \mu_0 \frac{\partial^2}{\partial t^2}\vec{P}^{\left(1\right)}\left(\vec{r},t\right) + \mu_0 \frac{\partial^2}{\partial t^2}\vec{P}^{\mathrm{NL}}\left(\vec{r},t\right) \,
\label{eq:genwaveeq}
\end{equation}
where $\epsilon_0$ and $\mu_0$ are respectively the permittivity and permeability of free-space in SI units, and $c = 1/\sqrt{\epsilon_0 \mu_0}$ is the speed-of-light in vacuum. The vector $\vec{E}$ represents the electric field at position $\vec{r}$ and at time $t$, and the vectors $\vec{P}^{\left(1\right)}$ and $\vec{P}^{\mathrm{NL}}$ are respectively the linear and nonlinear polarizations of the dielectric medium. Note that certain materials have non-negligible magnetic contributions to the nonlinear electric polarization response,\cite{dreye18.01} but those materials are not being considered. Also note that we are only considering the dipolar response, where higher-order multipole moments are being neglected.

The charge density is given by Gauss's law,
\begin{align}
\rho\left(\vec{r},t\right) &= \nabla \cdot \vec{D}\left(\vec{r},t\right) \nonumber \\
&=  \epsilon_0 \nabla \cdot \vec{E}\left(\vec{r},t\right) + \nabla \cdot \vec{P}^{\left(1\right)}\left(\vec{r},t\right) + \nabla \cdot \vec{P}^{\mathrm{NL}}\left(\vec{r},t\right) \, .
\label{eq:gauss}
\end{align}
It follows that
\begin{equation}
\epsilon_0 \nabla \cdot \vec{E}\left(\vec{r},t\right) = \rho\left(\vec{r},t\right) - \nabla \cdot \vec{P}^{\left(1\right)}\left(\vec{r},t\right) - \nabla \cdot \vec{P}^{\mathrm{NL}}\left(\vec{r},t\right) \, .
\label{eq:deldotE}
\end{equation}

From Eq. \ref{eq:genwaveeq} and \ref{eq:deldotE}, the wave equation for the propagation of light through a nonlinear dielectric medium can be rewritten as
\begin{align}
&\nabla^2 \vec{E}\left(\vec{r},t\right) - \frac{1}{\epsilon_0}\nabla \left[\rho\left(\vec{r},t\right) - \nabla \cdot \vec{P}^{\left(1\right)}\left(\vec{r},t\right) - \nabla \cdot \vec{P}^{\mathrm{NL}}\left(\vec{r},t\right)\right] = \frac{1}{c^2}\frac{\partial^2 \vec{E}\left(\vec{r},t\right)}{\partial t^2} \\
&+ \mu_0 \frac{\partial^2}{\partial t^2}\vec{P}^{\left(1\right)}\left(\vec{r},t\right) + \mu_0 \frac{\partial^2}{\partial t^2}\vec{P}^{\mathrm{NL}}\left(\vec{r},t\right) \nonumber \, .
\label{eq:genwaveeq2}
\end{align}

The linear polarization in the time domain is a convolution of the second-rank tensor, $\overline{\overline{X}}$, and the electric field,
\begin{equation}
\vec{P}^{\left(1\right)}\left(\vec{r},t\right) = \displaystyle \epsilon_0 \int_{-\infty}^{t} dt'\, \overline{\overline{X}}^{\left(1\right)}\left(\vec{r},t-t'\right) \cdot \vec{E} \left(\vec{r},t'\right) \, ,
\label{eq:vecPlin}
\end{equation}
In component form, the time-dependent linear polarization is given by
\begin{equation}
P_{\alpha}^{\left(1\right)}\left(\vec{r},t\right) = \displaystyle \epsilon_0 \int_{-\infty}^{t} dt' \, X_{\alpha\beta}^{\left(1\right)}\left(\vec{r},t-t'\right) E_\beta \left(\vec{r},t'\right) \, .
\label{eq:cartPlin}
\end{equation}
The Greek subscripts represent Cartesian coordinates in Eq. \ref{eq:cartPlin}, and there are no distinctions made between covariant and contravariant tensor components. The nonlinear polarization is expressed as a series, which is expanded in powers of the electric field. In Cartesian coordinates, the $\alpha$ component of the nonlinear polarization follows as
\begin{align}
\label{eq:cartPNL}
P_{\alpha}^{\mathrm{NL}} \left(\vec{r},t\right) &= \epsilon_0 \displaystyle \int_{-\infty}^{t}\int_{-\infty}^{t} dt_1\,dt_2\, X_{\alpha\beta\gamma}^{\left(2\right)}\left(\vec{r},t-t_1,t-t_2\right) E_\beta \left(\vec{r},t_1\right) E_\gamma \left(\vec{r},t_2\right) \\
&+ \displaystyle \epsilon_0 \int_{-\infty}^{t}\int_{-\infty}^{t}\int_{-\infty}^{t} dt_1\,dt_2\,dt_3 \, X_{\alpha\beta\gamma\delta}^{\left(3\right)}\left(\vec{r},t-t_1,t-t_2,t-t_3\right) \nonumber \\
&\times E_\beta \left(\vec{r},t_1\right) E_\gamma \left(\vec{r},t_2\right) E_\delta \left(\vec{r},t_3\right) + \cdots \nonumber \, .
\end{align}

The current set of equations in the time domain is quite difficult to handle analytically due to the presence of nonlinear convolutions. A constant amplitude and sinusoidal function is one possible solution to the linear wave equation, which describes a monochromatic plane wave. When the local response function is static over time, e.g. no changes in a homogeneous material due to heating, reorientation, chemical reactions, etc., then the wave equation for light propagating in a nonlinear medium (including all electric dipole polarization response functions) can be written in the frequency domain. In the time domain, more complex optical waveforms can be created in nature; however, these waves can be constructed from a set of sinusoidal waves. The Fourier transform of the time domain to the frequency domain is defined as
\begin{equation}
\vec{E}(\vec{r},\omega) =\displaystyle \int_{-\infty}^{\infty} \vec{E}(\vec{r},t)  e^{j\omega t} dt \, .
\label{eq:fourier}
\end{equation}
Because each frequency component is independent, the linear polarization in the frequency domain is simply given by
\begin{equation}
P_{\alpha}^{\left(1\right)}\left(\vec{r},\omega\right) = \epsilon_0 \chi_{\alpha\beta}^{\left(1\right)}\left(\vec{r},\omega;\omega\right)E_\beta \left(\vec{r},\omega\right) \, .
\label{eq:chiPlin}
\end{equation}
The linear response function in this form, $\chi^{\left(1\right)}$, is commonly referred to as the linear electric susceptibility. The nonlinear polarization in the frequency domain follows as
\begin{align}
\label{eq:chiPNL}
P_{\alpha}^{\mathrm{NL}}\left(\vec{r},\omega_m\right) &= \displaystyle \epsilon_0 \sum_{\beta\gamma} \sum_{lk} \chi_{\alpha\beta\gamma}^{\left(2\right)} \left(\vec{r},\omega_m = \omega_l + \omega_k ; \omega_l,\omega_k\right) E_\beta \left(\vec{r},\omega_l\right) E_\gamma \left(\vec{r},\omega_k\right) \\
&+ \displaystyle \epsilon_0 \sum_{\beta\gamma\delta} \sum_{lku} \chi_{\alpha\beta\gamma\delta}^{\left(3\right)} \left(\vec{r},\omega_m = \omega_l + \omega_k + \omega_u ; \omega_l,\omega_k,\omega_u\right) \nonumber \\
&\times E_\beta \left(\vec{r},\omega_l\right) E_\gamma \left(\vec{r},\omega_k\right) E_\delta \left(\vec{r},\omega_u\right) + \cdots \nonumber \, .
\end{align}
where all frequencies can be positive or negative. It is obvious why most analytical nonlinear optical calculations are performed in the frequency domain, where the lack of multiple time integrals can significantly reduce the complexity of problems for a discrete number of frequencies. Note that there are several properties of the nonlinear susceptibility tensor that can be used to relate the elements, and thereby reduce the total number of independent parameters.

\section{Review of Simplified Second-Order Frequency Mixing}

The three-dimensional wave equation for an electric field in a second-order nonlinear optical material in vector-component form is given by
\begin{align}
\label{eq:waveeqcompform}
&\displaystyle \sum_m \sum_\alpha \frac{\partial}{\partial r_\alpha} \left[\frac{\partial}{\partial r_\alpha} \sum_{\beta} E_{\beta,m}\left(\vec{r},\omega_m \right) \hat{r}_\alpha - \sum_{\beta} \frac{\partial}{\partial r_\beta} E_{\beta,m}\left(\vec{r},\omega_m \right) \hat{r}_\beta \right] = \\
&-\sum_m \frac{\omega_{m}^2}{c^2}\sum_\alpha \Bigg[ E_{\alpha,m}\left(\vec{r},\omega_m \right) + \chi_{\alpha\beta}^{\left(1\right)}\left(\vec{r},\omega_m;\omega_m\right)E_{\beta,m} \left(\vec{r},\omega_m\right) \nonumber \\
&+ \displaystyle \sum_{\beta\gamma} \sum_{lk} \chi_{\alpha\beta\gamma}^{\left(2\right)} \left(\vec{r},\omega_m = \omega_l + \omega_k ; \omega_l,\omega_k\right) E_{\beta,l} \left(\vec{r},\omega_l\right) E_{\gamma,k} \left(\vec{r},\omega_k\right)\Bigg]\hat{r}_\alpha \nonumber \, ,
\end{align}
where $r_\alpha$ denotes the Cartesian coordinates $x$, $y$, and $z$. Likewise, $\hat{r}_\alpha$ refers to the Cartesian unit vector. Note that we have already assumed a traveling wave solution based on the form of Eq. \ref{eq:waveeqcompform} implemented as a series of separate terms with different frequencies.

Equation \ref{eq:waveeqcompform} can lead to quite complicated analytical expressions. Thus, we will simplify second-order frequency mixing problems by assuming a plane wave in an infinite nonlinear optical medium. The nonlinear medium also is assumed to contain no free charges. To further simplify, let us assume that the electric field is linearly polarized and the wave is propagating in the direction of the positive $x$ axis. We also assume that the second-order nonlinear material is homogeneous. Let us further assume that the Kleinman symmetry condition holds, where the scalar wave equations can be rewritten using the convention $\chi^{2} = 2 d_\mathrm{eff}$. Under these approximations, Eq. \ref{eq:waveeqcompform} reduces to a simplified form,
\begin{align}
\label{eq:simpleNLOeq}
&\displaystyle \sum_m \frac{d^2}{d x^2} E_m\left(x,\omega_m \right) = -\sum_m \frac{\omega_{m}^2}{c^2} \Bigg[E_m\left(x,\omega_m \right) \\
&+ \chi^{\left(1\right)}\left(\omega_m;\omega_m\right)E_m\left(x,\omega_m\right) + \displaystyle 2 \sum_{lk} d_\mathrm{eff} E_l\left(x,\omega_l\right) E_k\left(x,\omega_k\right)\Bigg] \nonumber\, .
\end{align}
Note that $E_m$ and $\chi^{\left(1\right)}$ are now frequency dependent scalars. The amplitudes of the plane waves for each frequency component are known at $x = 0$ immediately after they enter the nonlinear material.

We are interested in finding approximate solutions to Eq. \ref{eq:simpleNLOeq} which have linear solutions oscillating at individual frequencies, each an orthogonal oscillating function, and with varying amplitudes. Thus, we can rewrite Eq. \ref{eq:simpleNLOeq} as
\begin{align}
& \displaystyle \sum_m \frac{d^2}{d x^2} A_m\left(x\right)\cos\left(n_m \omega_m x/c - \omega_m t\right) = -\sum_m \frac{\omega_{m}^2}{c^2} \Bigg[1 + \chi^{\left(1\right)}\left(\omega_m;\omega_m\right)\Bigg] A_m\left(x\right)\cos\left(n_m \omega_m x/c - \omega_m t\right) \nonumber \\
&+ \displaystyle 4 \sum_m \sum_{lk} d_\mathrm{eff} A_l\left(x\right)A_k\left(x\right)\cos\left(n_l \omega_l x/c - \omega_l t\right) \cos\left(n_k \omega_k x/c - \omega_k t\right) \, .
\label{eq:NLOamp1}
\end{align}
We have explicitly written the form of the assumed solution of an oscillating wave with a slow-varying amplitude,
\begin{equation}
E_m\left(x,\omega_m\right) =2A_m\left(x\right)\cos\left(n_m \omega_m x/c - \omega_m t\right) \, ,
\label{eq:Emslowvar}
\end{equation}
where
\begin{equation}
n_m = \sqrt{1+\chi^{\left(1\right)}\left(\omega_m;\omega_m\right)} \, .
\label{eq:Kndef}
\end{equation}
The cosine function can be rewritten using Euler's formula. Equation \ref{eq:NLOamp1} can then be rewritten as
\begin{align}
& \displaystyle \sum_m \frac{d^2}{d x^2} \Bigg[A_m\left(x\right)e^{j\left(n_m \omega_m x/c - \omega_m t\right)} + \mathrm{c.\,c.}\Bigg] = \sum_m n_{m}^2 \frac{\omega_{m}^2}{c^2} \Bigg[A_m\left(x\right)e^{j\left(n_m \omega_m x/c - \omega_m t\right)} + \mathrm{c.\,c.}\Bigg]  \nonumber \\
&+ \displaystyle 2 \frac{\omega_{m}^2}{c^2} \sum_m \sum_{lk} d_\mathrm{eff} \Bigg[A_l\left(x\right)e^{j\left(n_l \omega_l x/c - \omega_l t\right)} + \mathrm{c.\,c.}\Bigg] \Bigg[A_k\left(x\right)e^{j\left(n_k \omega_k x/c - \omega_k t\right)} + \mathrm{c.\,c.}\Bigg] \, ,
\label{eq:NLOamp2}
\end{align}
where $\mathrm{c.\,c.}$ denotes the complex conjugate of the left-hand terms in each bracket containing the symbol.

Because of the orthogonality condition for frequency components,
\begin{equation}
\displaystyle \int_{-\infty}^{\infty} e^{j\omega t} e^{-j\omega' t} \, dt = \delta\left(\omega' - \omega\right) \, ,
\label{eq:orthocond}
\end{equation}
Eq. \ref{eq:NLOamp2} may be rewritten as separate equations, where $\omega_m = \omega_1, \omega_2, \omega_3, \ldots$. Let us limit the study to three possible waves traveling at angular frequencies $\omega_1$, $\omega_2$, and $\omega_3$, where $\omega_1 + \omega_2 - \omega_3 = 0$. For such cases, the only nonlinear scenario, ignoring higher-order nonlinearities from microscopic cascading effects,\cite{baev10.01,dawso13.01} occurs when two of the frequencies either add or subtract and results in the third possible frequency. Including all relevant frequency mixing terms in the nonlinear interaction summation, the three wave equations follow as
\begin{align}
& \displaystyle \frac{d^2}{d x^2} \Bigg[A_1\left(x\right)e^{j\left(n_1 \omega_1 x/c - \omega_1 t\right)} + \mathrm{c.\,c.}\Bigg] = -n_{1}^2\frac{\omega_{1}^2}{c^2} \Bigg[A_1\left(x\right)e^{j\left(n_1 \omega_1 x/c - \omega_1 t\right)} + \mathrm{c.\,c.}\Bigg]  \nonumber \\
&- \displaystyle 4\frac{\omega_{1}^2}{c^2} d_\mathrm{eff} \Bigg[A_{2}^\ast \left(x\right) A_3 \left(x\right) e^{j\left[\left(n_3 \omega_3 - n_2 \omega_2\right) x/c - \omega_1 t\right]} + \mathrm{c.\,c.}\Bigg] \, ,
\label{eq:w1} \\
& \displaystyle \frac{d^2}{d x^2} \Bigg[A_2\left(x\right)e^{j\left(n_2 \omega_2 x/c - \omega_2 t\right)} + \mathrm{c.\,c.}\Bigg] = -n_{2}^2\frac{\omega_{2}^2}{c^2} \Bigg[A_2\left(x\right)e^{j\left(n_2 \omega_2 x/c - \omega_2 t\right)} + \mathrm{c.\,c.}\Bigg]  \nonumber \\
&- \displaystyle 4\frac{\omega_{2}^2}{c^2} d_\mathrm{eff} \Bigg[A_{1}^\ast \left(x\right) A_3 \left(x\right)e^{j\left[\left(n_3 \omega_3 - n_1 \omega_1\right) x/c - \omega_2 t\right]} + \mathrm{c.\,c.}\Bigg] \, ,
\label{eq:w2} \\
& \displaystyle \frac{d^2}{d x^2} \Bigg[A_3\left(x\right)e^{j\left(n_3 \omega_3 x/c - \omega_3 t\right)} + \mathrm{c.\,c.}\Bigg] = -n_{3}^2 \frac{\omega_{3}^2}{c^2} \Bigg[A_3\left(x\right)e^{j\left(n_3 \omega_3 x/c - \omega_3 t\right)} + \mathrm{c.\,c.}\Bigg]  \nonumber \\
&- \displaystyle 4\frac{\omega_{3}^2}{c^2} d_\mathrm{eff} \Bigg[A_1\left(x\right) A_2 \left(x\right)e^{j\left[\left(n_1 \omega_1 + n_2 \omega_2\right) x/c - \omega_3 t\right]} + \mathrm{c.\,c.}\Bigg] \, .
\label{eq:w3}
\end{align}
Due to the symmetry of real valued oscillating functions expressed as a clockwise and a counter-clockwise motion oscillating at the same frequency on the complex unit circle, either the explicitly given terms in Eqs. \ref{eq:w1}-\ref{eq:w3} or the complex conjugates will alone satisfy the equalities. Therefore, without loss of generality, we can subtract the complex conjugate terms from both sides of Eqs. \ref{eq:w1}-\ref{eq:w3} leaving a complex amplitude equation. Afterward, the time dependence can be divided out of the equations.

The assumed form of the solution with a position dependent amplitude multiplied by a complex oscillating function allows us to use the product rule,
\begin{equation}
\displaystyle \frac{d^2}{dx^2}A_m\left(x\right) e^{j n_m \omega_m x/c} = e^{j n_m \omega_m x/c}\Bigg[\frac{d^2}{dx^2}A_m\left(x\right) + 2 j n_m \frac{\omega_m}{c} \frac{d}{dx}A_m\left(x\right)  - n_{m}^2 \frac{\omega_{m}^2}{c^2} A_m\left(x\right)\Bigg] \, .
\label{eq:chainchain}
\end{equation}
Thus, we may rewrite Eqs. \ref{eq:w1}-\ref{eq:w3} using Eq. \ref{eq:chainchain} and then divide by $e^{j n_m\omega_m x/c}$,
\begin{align}
& \displaystyle \left[\frac{d^2}{d x^2} + 2j n_{1}\frac{\omega_1}{c} \frac{d}{dx}\right]A_1\left(x\right) = -\displaystyle 4\frac{\omega_{1}^2}{c^2} d_\mathrm{eff} A_{2}^\ast\left(x\right) A_3 \left(x\right) e^{j\left(n_3 \omega_3 - n_2 \omega_2 - n_1 \omega_1\right) x/c} \, ,
\label{eq:w1amppre} \\
& \displaystyle \left[\frac{d^2}{d x^2} + 2j n_{2}\frac{\omega_2}{c} \frac{d}{dx}\right]A_2\left(x\right) = -\displaystyle 4\frac{\omega_{2}^2}{c^2} d_\mathrm{eff} A_{1}^\ast\left(x\right) A_3 \left(x\right) e^{j\left(n_3 \omega_3 - n_2 \omega_2 - n_1 \omega_1\right) x/c} \, ,
\label{eq:w2amppre} \\
& \displaystyle \left[\frac{d^2}{d x^2} + 2j n_{3}\frac{\omega_3}{c} \frac{d}{dx}\right]A_3\left(x\right) = -\displaystyle 4\frac{\omega_{3}^2}{c^2} d_\mathrm{eff} A_1\left(x\right) A_2 \left(x\right) e^{j\left(n_1 \omega_1 + n_2 \omega_2 - n_3 \omega_3\right) x/c} \, .
\label{eq:w3amppre}
\end{align}

The slow-varying amplitude approximation can be made when the following condition holds,
\begin{equation}
\left|\frac{d^2 A_m}{dx^2}\right| \ll \left|\frac{\omega_m}{c}\frac{d A_m}{dx}\right| \, .
\label{eq:slowvaryapprox}
\end{equation}
Setting $dA_{m}^2/dx^2 \approx 0$ in Eqs. \ref{eq:w1amppre}-\ref{eq:w3amppre} results in the final simplified expressions for three-wave mixing,
\begin{align}
\displaystyle \frac{d A_1}{dx} &= 2j\displaystyle \frac{\omega_{1}}{n_1 c} d_\mathrm{eff} A_{2}^\ast A_3 e^{j\left(n_3 \omega_3 - n_2 \omega_2 - n_1 \omega_1\right) x/c} \, , \label{eq:w1amp} \\
\displaystyle \frac{d A_2}{dx} &= 2j\displaystyle \frac{\omega_{2}}{n_2 c} d_\mathrm{eff} A_{1}^\ast A_3  e^{j\left(n_3 \omega_3 - n_2 \omega_2 - n_1 \omega_1\right) x/c} \, , \label{eq:w2amp} \\
\displaystyle \frac{d A_3}{dx} &= 2j\displaystyle \frac{\omega_{3}}{n_3 c} d_\mathrm{eff} A_{1} A_2 e^{j\left(n_1 \omega_1 + n_2 \omega_2 - n_3 \omega_3\right) x/c} \, . \label{eq:w3amp}
\end{align}
Equations \ref{eq:w1amp}-\ref{eq:w3amp} are a set of three interacting nonlinear equations. The spatial dependent amplitudes can be multiplied by their respective oscillating wave functions to give approximate solutions to Eqs. \ref{eq:w1}-\ref{eq:w3}.

\section{Homotopy Analysis Method Applied to Second-Order Wave Mixing}

The following describes the basic idea of HAM. Let
\begin{align}
\mathcal{N}_1[A_1(x)] &=
\displaystyle\frac{d A_1}{dx} - 2j \frac{\omega_{1}}{n_1 c} d_\mathrm{eff} A_{2}^\ast A_3 e^{j\left(n_3 \omega_3 - n_2 \omega_2 - n_1 \omega_1\right) x/c} \, ,
\label{nolinear1} \\
\mathcal{N}_2[A_2(x)] &=
\displaystyle \frac{d A_2}{dx} - 2j \frac{\omega_{2}}{n_2 c} d_\mathrm{eff} A_{1}^\ast A_3  e^{j\left(n_3 \omega_3 - n_2 \omega_2 - n_1 \omega_1\right) x/c} \, ,
\label{nolinear2} \\
\mathcal{N}_3[A_3(x)] &=
\displaystyle \frac{d A_3}{dx} - 2j \frac{\omega_{3}}{n_3 c} d_\mathrm{eff} A_{1} A_2 e^{j\left(n_1 \omega_1 + n_2 \omega_2 - n_3 \omega_3\right) x/c} \, ,
\label{nolinear3} \\
\end{align}
where $\mathcal{N}_1[A_1(x)] = \mathcal{N}_2[A_2(x)] = \mathcal{N}_3[A_3(x)] = 0$.

By using the technique of HAM \cite{Beyond}, we construct the zeroth-order deformation equations,
\begin{align}
(1-p)\mathcal{L}_1[A_1(x; p)-e_{10}(x)] &= ph \mathcal{H}_1(x)\mathcal{N}_1[A_1(x; p)] \, ,
\label{HP1} \\
(1-p)\mathcal{L}_2[A_2(x; p)-e_{20}(x)] &= ph \mathcal{H}_2(x)\mathcal{N}_2[A_2(x; p)] \, ,
\label{HP2} \\
(1-p)\mathcal{L}_3[A_3(x; p)-e_{30}(x)] &= ph \mathcal{H}_3(x)\mathcal{N}_3[A_3(x; p)] \, ,
\label{HP3}
\end{align}
where $p\in [0, 1]$ is the embedding parameter, $h \neq 0$ is an auxiliary parameter, and $\mathcal{L}_m$ are auxiliary linear operators. The $H_m(x)$ denote the nonzero auxiliary functions. The solution to each separate frequency-component wave equation will be of the form,
\begin{equation}
A_m(x;p)=e_{m0}\left(x\right)+\sum_{q=1}^{\infty}e_{mq}(x)p^q \, .
\end{equation}
For the linear operators,
\begin{equation}
\mathcal{L}_m(.)= \frac{\partial(.)}{\partial x} \, ,
\label{eq:lindif}
\end{equation}
we get
\begin{equation}
\mathcal{L}_m(.)^{-1}=\int^x (.) ds + b_{mq},\quad \mathcal{L}(b_{mq})=0 \, ,
\label{eq:lindifinv}
\end{equation}
where $b_{nq}$ is a constant of integration for the $q$th iteration of the $m$th equation.

We see when $p=0$ and $p=1$, $A_m(x; 0)=e_{mq}(x)$ and $A_m(x; 1)=A_m(x)$, which must be one of the solutions to a nonlinear equation $\mathcal{N}_m [A_m(x; p)]=0$ as proven by Liao.\cite{liao1,liao2} Expanding $A_m(x; p)$ in a Taylor series with respect to $p$,
\begin{equation}
\label{SR}
A_m(x; p)=e_{m0}(x)+\sum_{q=1}^{\infty}e_{mq}(x)p^q,\quad \mathrm{where}\quad e_{mq}(x)=\frac{1}{q!}\frac{\partial^q A_m(x; p)}{\partial p^q}\big|_{p=0}.
\end{equation}
We then define the vector,
\begin{equation}
\vec{e}_{mq}(x) =(e_{m0}(x),e_{m1}(x),e_{m2}(x),\ldots) \, .
\label{eq:vecemqeq}
\end{equation}

Differentiating the zeroth-order deformation equations, Eqs. \ref{HP1}-\ref{HP3}, $q$-times with respect to $p$, dividing them by $q!$, and then setting $p=0$, results in the $q$th-order deformation equations,
\begin{align}
\mathcal{L}_m[e_{mq}(x)-\xi_{q} e_{m\left(q-1\right)}(x)]=h \Re_{mq}\vec{e_{mq}}(x) \, ,
\label{Def}
\end{align}
where
\begin{equation}
\xi_{q}=
\begin{cases}
0,\quad q\leq 1\\
1,\quad q>1 \, .
\end{cases}
\label{eq:xiequation}
\end{equation}
and
\begin{equation}
\Re_{mq}(\vec{e}_{mq}(x))=\frac{1}{q!}\frac{\partial^{q-1} \mathcal{N}_m[A_m(x;p)]}{\partial p^{q-1}}\big|_{p=0} \, .
\label{eq:Renq}
\end{equation}
Substituting Eqs. \ref{nolinear1}-\ref{nolinear3} into Eq. \ref{eq:Renq}, we find
\begin{align}
\label{eq:Re1}
\Re_{1q}(\vec{e}_{1q}(x)) &= \frac{d}{dx} e_{1\left(q-1\right)}(x) \\
&- 2j \frac{\omega_{1}}{n_1 c}\sum_{u=0}^{q-1} d_\mathrm{eff} e_{2u}^\ast(x)e_{3\left(q-u-1\right)}(x) e^{j\left(n_3 \omega_3 - n_2 \omega_2 - n_1 \omega_1\right) x/c} \, , \nonumber \\
\label{eq:Re2}
\Re_{2q}(\vec{e}_{2q}(x)) &= \frac{d}{dx} e_{2\left(q-1\right)}(x) \\
&- 2j \frac{\omega_{2}}{n_2 c}\sum_{u=0}^{q-1} d_\mathrm{eff} e_{1u}^\ast(x)e_{3\left(q-u-1\right)}(x) e^{j\left(n_3 \omega_3 - n_2 \omega_2 - n_1 \omega_1\right) x/c} \, , \nonumber \\
\label{eq:Re3}
\Re_{3q}(\vec{e}_{3q}(x)) &= \frac{d}{dx} e_{3\left(q-1\right)}(x) \\
&- 2j \frac{\omega_{3}}{n_3 c}\sum_{u=0}^{q-1} d_\mathrm{eff} e_{1u}(x)e_{2\left(q-u-1\right)}(x) e^{j\left(n_1 \omega_1 + n_2 \omega_2 - n_3 \omega_3\right) x/c} \, . \nonumber
\end{align}
Following Eq. \ref{Def}, we can now write expressions for iteratively determining the $e_{nq}$ terms,
\begin{align}
\label{eq:long1}
e_{1q}(x)&=\xi_q  e_{1\left(q-1\right)}(x)+ h \int_{0}^x \frac{d}{ds} e_{1\left(q-1\right)}(s)\,ds \\
&- 2j h\frac{\omega_{1}}{n_1 c} d_\mathrm{eff} \sum_{u=0}^{q-1}\int_{0}^x e_{2u}^\ast(s)e_{3\left(q-u-1\right)}(s) e^{j\left(n_3 \omega_3 - n_2 \omega_2 - n_1 \omega_1\right) s/c} ds \, , \nonumber \\
\label{eq:long2}
e_{2q}(x)&=\xi_q  e_{2\left(q-1\right)}(x)+ h \int_{0}^x \frac{d}{ds} e_{2\left(q-1\right)}(s)\,ds \\
&- 2j h\frac{\omega_{2}}{n_2 c} d_\mathrm{eff} \sum_{u=0}^{q-1}\int_{0}^x e_{1u}^\ast(s)e_{3\left(q-u-1\right)}(s) e^{j\left(n_3 \omega_3 - n_2 \omega_2 - n_1 \omega_1\right) s/c} ds \, , \nonumber \\
\label{eq:long3}
e_{3q}(x)&=\xi_q  e_{3\left(q-1\right)}(x)+ h \int_{0}^x \frac{d}{ds} e_{3\left(q-1\right)}(s)\,ds \\
&- 2j h\frac{\omega_{3}}{n_3 c} d_\mathrm{eff} \sum_{u=0}^{q-1}\int_{0}^x e_{1u}(s)e_{2\left(q-u-1\right)}(s) e^{j\left(n_1 \omega_1 + n_2 \omega_2 - n_3 \omega_3\right) s/c} ds \, , \nonumber
\end{align}

\section{Results}

Consider the linear differential operators, $dA_{m}^\mathrm{lin}/dx = 0$. The solutions to the three linear amplitude equations, ($A_{1}^\mathrm{lin},\,A_{2}^\mathrm{lin},\,A_{3}^\mathrm{lin}$), are all constants ($a_1,\,a_2,\,a_3$) determined by the left boundary value at $x = 0$, where the direction of propagation points to the right. In general, the constants $a_1$, $a_2$, and $a_3$ are complex amplitudes. The solutions to the linear differential operators are used as the initial guess in the HAM approach, \textit{i}.\textit{e}., $(e_{10},e_{20}e_{30}) = (a_1,\,a_2,\,a_3)$. Higher-order deformations are iteratively determined via Eqs. \ref{eq:long1}-\ref{eq:long3}, where the approximation is obtained after summing each term according to Eq. \ref{SR} out to the highest order of the truncated Taylor series and letting $q \rightarrow 1$.

\begin{center}
\begin{table}[t!]
\caption{Parameters for HAM comparison with numerical results.}
\centering
\begin{tabular}{c c c}
\hline
Parameter & value & units \\ \colrule
$\omega_1/2\pi$ & 250 & THz \\
$\omega_2/2\pi$ & 350 & THz \\
$\omega_3/2\pi$ & 600 & THz \\
$P_1$ & 100 & MW \\
$P_2$ & 50 & MW \\
$P_3$ & 20 & kW \\
$R$ & 2.5 & mm \\
$d_\mathrm{eff}$ & 2 & pm/V \\ \botrule
\hline
\end{tabular}
\label{table:1}
\end{table}
\end{center}

\begin{figure}[t!]
\centering\includegraphics[scale=1]{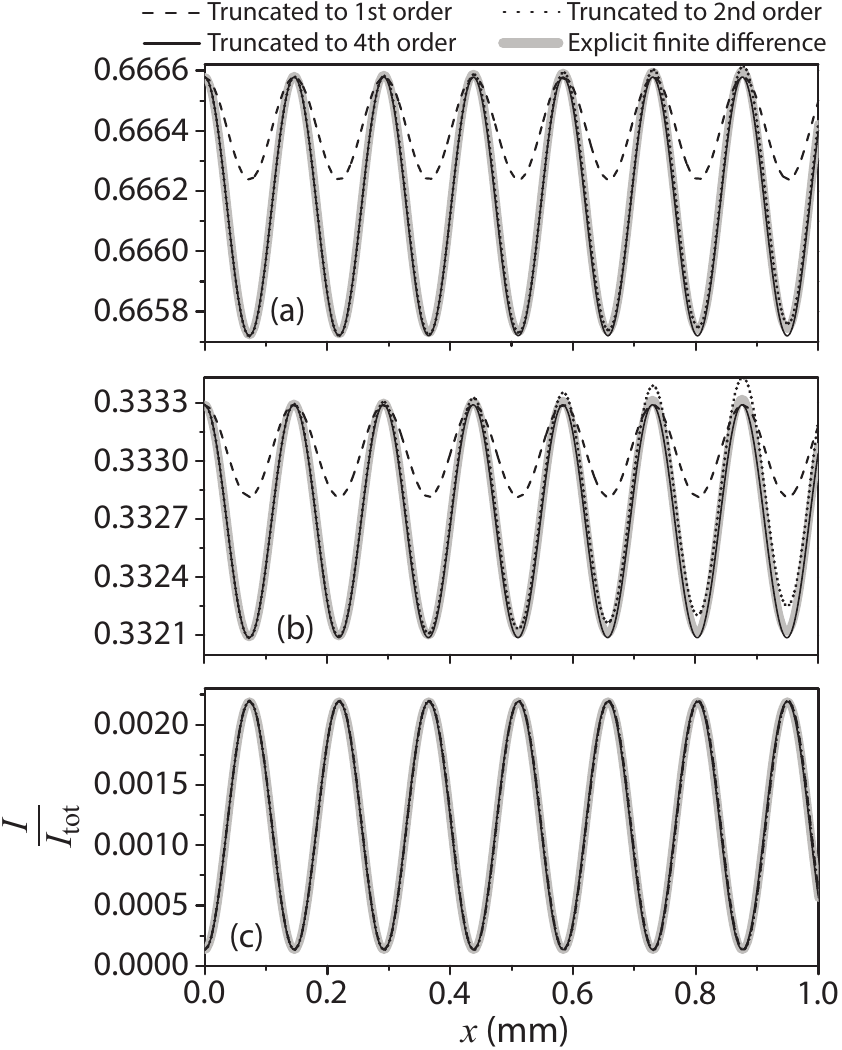}
\caption{Phase mismatched results with $n_1 = 1.776$, $n_2 = 1.777$, and $n_3 = 1.780$ for the normalized intensities of waves oscillating at (a) $\omega_1$, (b) $\omega_2$, and (c) $\omega_3$. The HAM approximation with $h = -1$ out to $4$th order from an initial guess determined by the linear differential operator is compared to an explicit finite-difference approximation. The parameters are given in Table I.}
\label{fig:mismatpoly}
\end{figure}

A phase mismatch can occur in nonlinear dispersive media, where it is convenient to define the difference in wave numbers,
\begin{equation}
\Delta k = \left(n_3 \omega_3 - n_2 \omega_2 - n_1 \omega_1\right)/c \, .
\label{eq:phase}
\end{equation}
When $\Delta k \neq 0$, the mismatch in phase over distances causes the nonlinear mixing to be generated and quickly depleted over short cycle governed by $\Delta k$. For small values of $\Delta k$ the oscillations in phase mismatched generation/depletion are slow, where increasing $\Delta k$ quickly results in oscillations that are much faster and smaller in amplitude.

The first few orders in the HAM approximation for a phase mismatched scenario are given in \ref{app:mismatch}. The HAM results are plotted in Fig. \ref{fig:mismatpoly} for $h = -1$ and compared to numerical results. The field amplitudes at the left boundary were determined by first assuming a plane wave with a power $P$ measured over a small circular area of radius $R$, where the intensity is given by
\begin{equation}
I = \frac{P}{\pi R^2} \, .
\label{eq:powerarea}
\end{equation}
Assuming negligible third-order and higher contributions to the light-matter interaction, the intensity is related to the field magnitude by
\begin{equation}
I_m\left(x=0\right) = \frac{1}{2}c n_m \epsilon_0 \left|a_m\right|^2 \, .
\label{eq:intensitym}
\end{equation}
For simplicity, the amplitudes were assumed real with zero phase at the boundary.

The amplitude oscillations out to a $1\,$mm depth are shown in Fig. \ref{fig:mismatpoly} by plotting the normalized intensities as a function of $x$, where $I_\mathrm{tot} = I_1 + I_2 + I_3$. Several amplitude oscillations resulting from the phase mismatch are shown, where $n_1 = 1.776$, $n_2 = 1.777$, and $n_3 = 1.780$ which corresponds to a value of $\Delta k \approx 43\,$rad/mm for the frequencies provided in Table I. The value of $\Delta k$ results approximately 6.8 amplitude oscillations over the length of a millimeter. The exponential functions in the nonlinear terms quickly enter the HAM approximation after the first-order iteration. By the second iteration the HAM approximation for all three amplitudes closely matches the numerical results over a few amplitude oscillations. The fourth order HAM approximation further increases the accuracy. The HAM approximation is compared to numerical results obtained using an explicit finite-difference scheme.

\begin{figure}[t!]
\centering\includegraphics[scale=1]{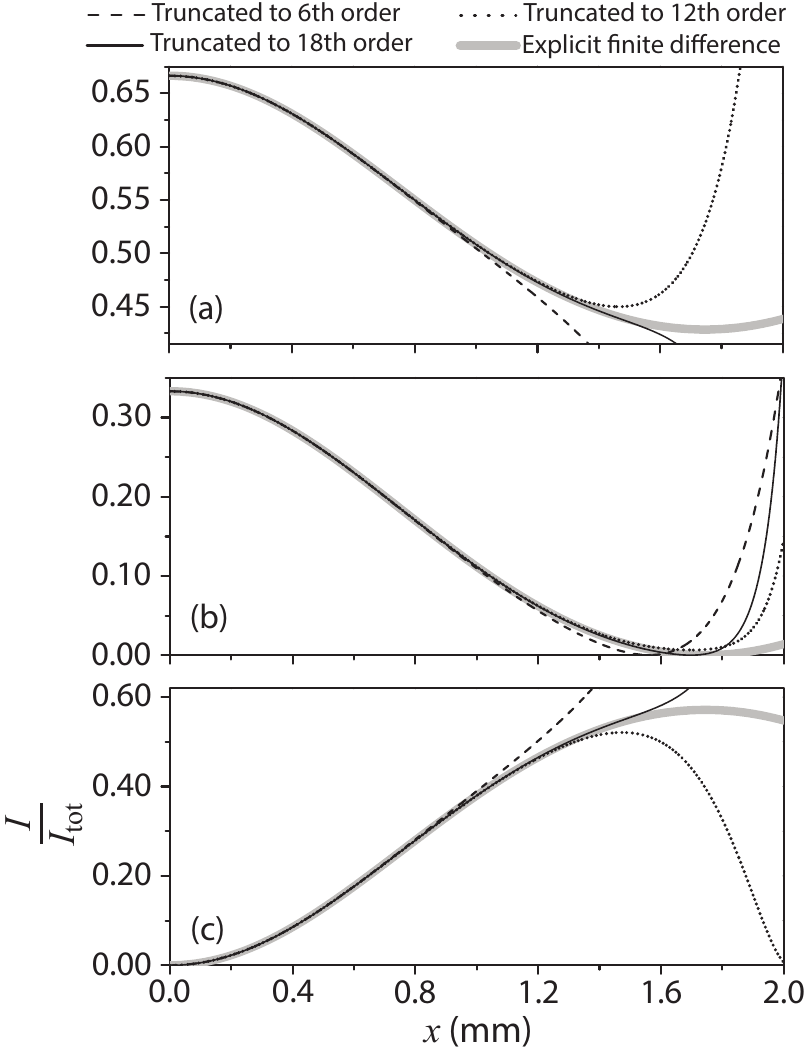}
\caption{The normalized intensities for the perfectly phased matched case with $n_1 = n_2 = n_3 = 1.78$ are plotted for fields oscillating at frequencies (a) $\omega_1$, (b) $\omega_2$, and (c) $\omega_3$. The HAM approximation out to $18$th order for $h = -1$ is compared to results from an explicit finite-difference scheme. The parameters are given in Table I, which corresponds to sum frequency generation at small $x$.}
\label{fig:PPMpoly}
\end{figure}

The \textsl{perfectly phase-matched} scenario occurs when $\Delta k = 0$. The most common experimental technique to obtain perfect phase matching utilizes the birefringence of anisotropic crystals, where an axis of a crystal is rotated out of plane to change the refractive indices of light polarized along specific directions.\cite{boyd09.01} The HAM approximation for the perfectly phase matched case, using the values given in Table I, is shown in Fig. \ref{fig:PPMpoly}. The normalized intensities corresponding to the frequencies $\omega_1, \, \omega_2, \, \omega_3$ are displayed as a function of the penetration depth through the nonlinear medium. The low-order iterations for $h = -1$ result in polynomial expressions which quickly converge to the numerical approximation up to the first inflection point. The HAM approximation expressed out to an 18th-order polynomial does not converge out to the first extremum for $x>0$ when $h = -1$ as illustrated in Fig. \ref{fig:PPMpoly}. Due to the fast convergence of the approximation out to the first inflection point, after only a few iterations the position of the inflection point can be determined via $dA^2/dx^2 = 0$. The inflection point is midway between local extrema, where determining the amplitude at the inflection point will determine the amplitude at the next extremum for a lossless medium. The periodicity of the solution for the case of a lossless medium allows for the amplitude to be approximated to the left and right at each extrema, which can be used to piece together the oscillating function if amplitudes need to be determined over a greater penetration depth.

The terms obtained from HAM for the perfectly phase-matched case are given in \ref{app:match}. The term $e_{mq}$ is a $q$th order polynomial. When $h = -1$, the term $e_{mq}$ becomes a power function to the $q$th power. There is no power mixing between terms for the power series basis used in our formulation of the HAM approximation to three-wave mixing when $h = -1$. The auxiliary parameter can be in the range $-2<h<0$, where the value affects the convergence region for $x$ as well as the accuracy of the function. As the auxiliary parameter is increased to a smaller negative number, the convergence region for $x$ increases. Clearly, if we want the series to be convergent for $0\leq x <\infty$, then $h$ would tend to zero. When $h\rightarrow 0$, then the approximation approaches the constant initial guess, which is convergent for $x$ out to infinity, but it is also a terrible approximation for finite values of $x$. For a polynomial representation of an oscillating function in general, we find that as the convergence region of $x$ is increased, the approximation becomes worse at small $x$. Therefore, there is an optimal value of $h$ over the defined region in which the field amplitudes are to be approximated from a truncated series solution.

\begin{figure}[t!]
\centering\includegraphics[scale=1]{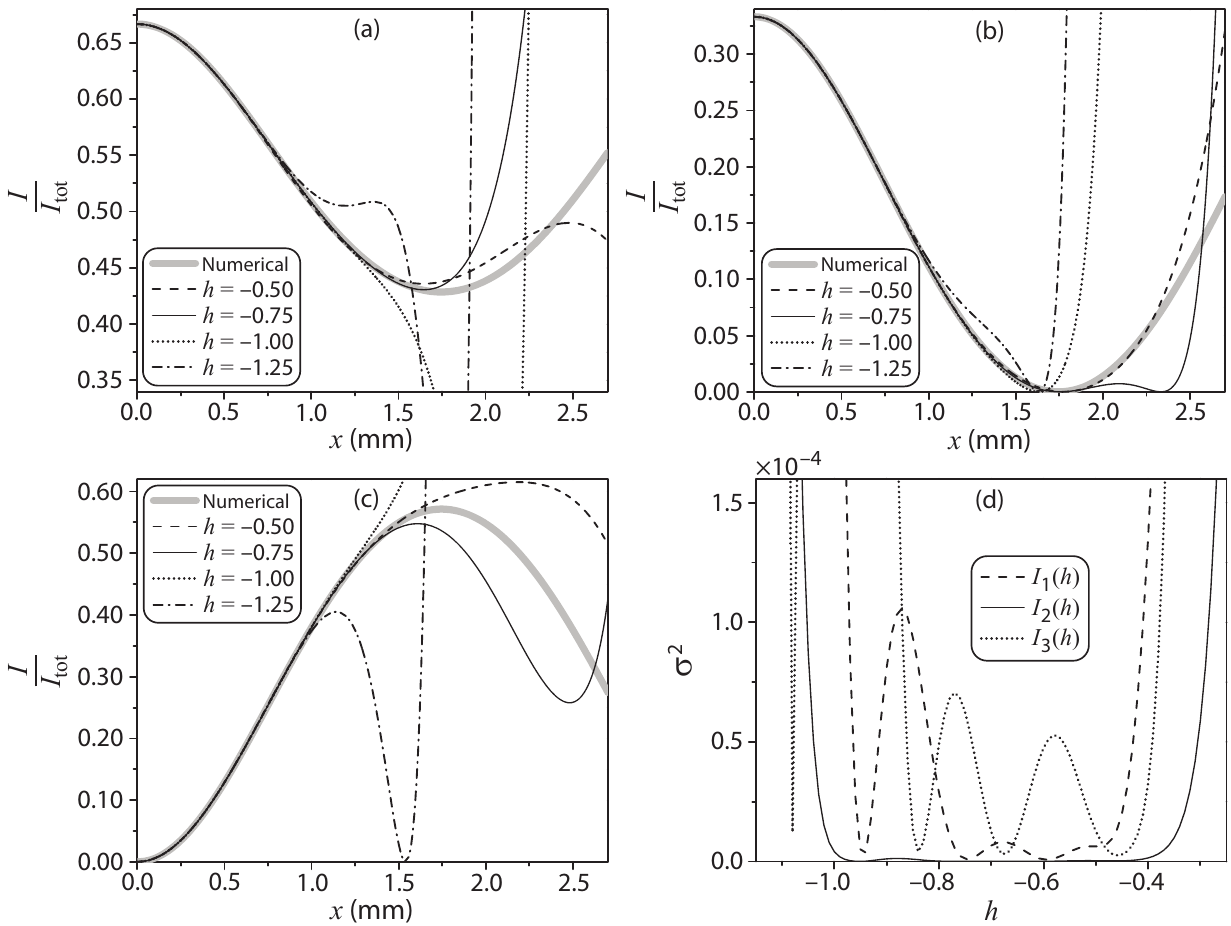}
\caption{Using the parameters in Table I, the normalized intensities for frequencies (a) $\omega_1$, (b) $\omega_2$, and (c) $\omega_3$ are plotted for the perfectly phased matched case with $n_1 = n_2 = n_3 = 1.78$. The $10$th order HAM approximation is shown for different values of the auxiliary parameter and compared to the numerical results. (d) The variance between the $10$th order HAM approximation and numerical results in the range between $x = 0$ and the first local minimum as a function of the auxiliary parameter.}
\label{fig:inth}
\end{figure}

The normalized intensities corresponding to the waves propagating with frequencies $\omega_1$, $\omega_2$, and $\omega_3$ are shown in Fig. \ref{fig:inth}(a)-(c), which have been approximated with HAM to $10$th order. The approximations are compared to results obtained by an explicit finite-difference scheme. The HAM results increase their convergence region in $x$ when $h$ is increased, where the $10$th-order HAM approximation more closely approximates the first extremum for $x>0$. Before the first inflection point, the $10$th-order HAM results are already very good approximations. Decreasing the auxiliary parameter below negative one has little benefit and can significantly decrease the convergence region as illustrated for $h = -1.25$. To see how the auxiliary parameter changes the convergence region and the goodness of the approximation, Fig. \ref{fig:inth}(d) shows the normalized intensity variance $\sigma^2$ between the extremum at $x = 0$ and the first extremum for $x>0$ as a function of the auxiliary parameter. The optimal auxiliary parameter for a specified range can be determined by minimizing the sum of each amplitude's variance.

\begin{figure}[t!]
\centering\includegraphics[scale=1]{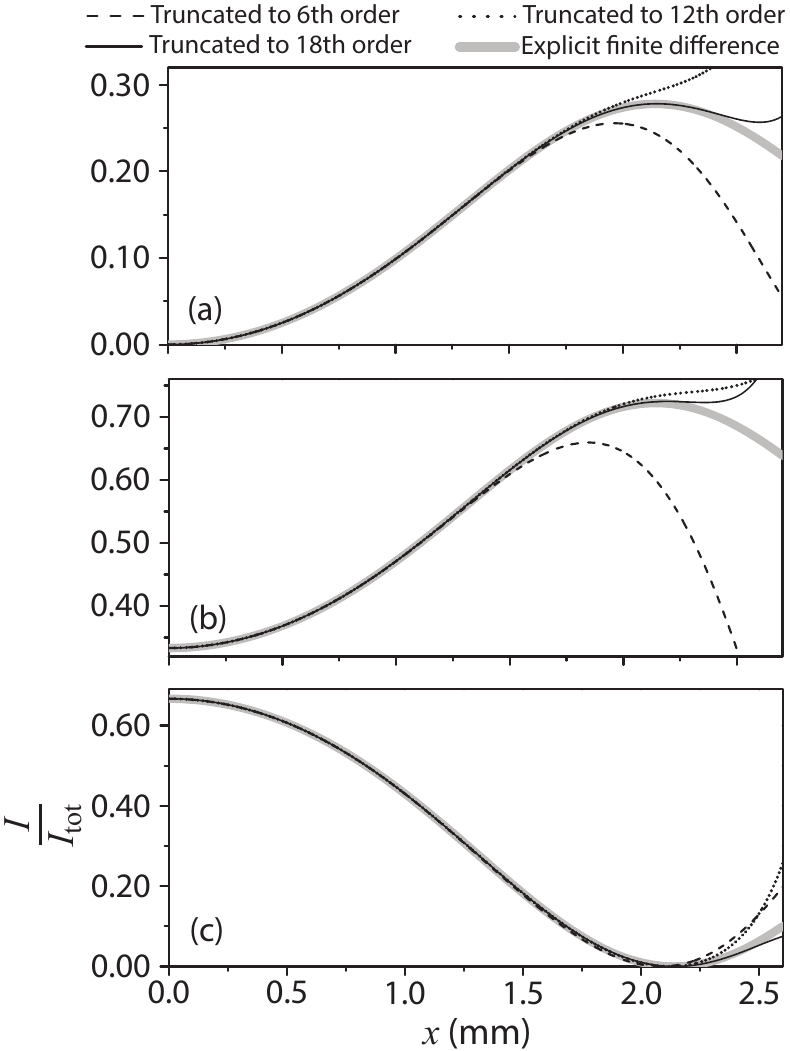}
\caption{The normalized intensities for frequencies (a) $\omega_1$, (b) $\omega_2$, and (c) $\omega_3$ are plotted for the perfectly phased matched case with $n_1 = n_2 = n_3 = 1.78$. The HAM approximation out to $18$th order for $h = -1$ is compared to results from an explicit finite-difference scheme. The parameters are close to those given in Table I, except that the field magnitudes $\left|a_1\right|$ and $\left|a_3\right|$ have been swapped which corresponds to the case of difference frequency generation at small $x$.}
\label{fig:FDppm}
\end{figure}

The HAM expression is an analytical approximation to the given system of equations. Unlike the exact analytical solution given by Armstrong et al. which requires the ranking of the roots of a cubic equation, there is no need to specify conditions beyond the boundary conditions. The HAM expressions work for any general three-wave mixing scenario involving a $\chi^{\left(2\right)}$ process. Using the parameters in Table I, but switching the powers measured over a small area for the waves traveling with frequencies $\omega_1$ and $\omega_3$, we arrive at a general difference frequency case with $P_1 = 20\,$kW and $P_3 = 100\,$MW. Using the exact same analytical expressions obtained from the HAM, the case corresponding to the seeded generation of light at a frequency corresponding to the difference in frequency of two other light waves is plotted along with the numerical results in Fig. \ref{fig:FDppm}. Again, low-order truncations for $h = -1$ well approximate the phenomenon beyond the inflection point. When $h = -1$, both the $12$th and $18$th order approximations capture the frequency mixing behavior to nearly the first extremum for $x > 0$.

\section{Conclusion}

Truncated HAM approximations of three-wave mixing have been determined under the scalar field and slow-varying amplitude approximations. The HAM results using a power basis were compared to numerical approximations, where we observe good agreement to numerical results beyond the first inflection point in all cases. The convergence region was shown to increase with an increase in the auxiliary parameter, which decreased the variance measured between $x = 0$ and the first extremum for $x > 0$. The variance began to increase dramatically when the auxiliary parameter was increased above $-1/2$.

Analytical expressions allow for symbolic manipulation to determine limits, symmetries, etc. The HAM approximation to nonlinear optical phenomena could be a valuable tool to generate analytical approximations for many types of higher-order nonlinear optical phenomena, where generalized higher-order equations have not been solved analytically. HAM expressions can be generated that contain only common functions which are easy to manipulate. We have demonstrated the ability of HAM to generate analytical expressions that approximate complicated nonlinear optical behavior, where study of the method applied to more complex scenarios could provide valuable new insights into a broader class of observable phenomena.

\appendix

\section{Phase mismatched HAM terms}
\label{app:mismatch}

The HAM terms for the general phase mismatched case under the slow-varying approximation for three-wave mixing is shown out to $q = 3$. The initial guesses are given by the left boundary conditions for right-traveling waves,
\begin{align}
e_{10} &= a_1 \, , \label{eq:e10calc} \\
e_{20} &= a_2 \, , \label{eq:e20calc} \\
e_{30} &= a_3 \, . \label{eq:e30calc}
\end{align}
The first-order HAM deformations follow as
\begin{align}
e_{11} &= 2 h \frac{d_\mathrm{eff} \omega_1 a_{2}^\ast a_3}{n_1 c \, \Delta k} \left(1 - e^{i \Delta k\, x}\right) \, , \\
e_{21} &= 2 h \frac{d_\mathrm{eff} \omega_2 a_{1}^\ast a_3}{n_2 c \, \Delta k} \left(1 - e^{i \Delta k\, x}\right) \, , \\
e_{31} &= 2 h \frac{d_\mathrm{eff} \omega_3 a_{1} a_2}{n_3 c \, \Delta k} \left(e^{-i \Delta k\, x} - 1\right) \, .
\end{align}
The second-order HAM deformations are given by
\begin{align}
e_{12} &= \frac{h d_\mathrm{eff} \omega_1}{c^2 \left(\Delta k\right)^2 n_1 n_2 n_3} \Big\{n_2 a_{2}^\ast \Big[2 a_3 n_3 c \Delta k \left(1+h\right)\left(1 - e^{i \Delta k \, x}\right) \\
&+ 4 h a_1 a_2 d_\mathrm{eff} \omega_3 \left(e^{i \Delta k \, x} - i \Delta k\, x - 1\right)\Big] \nonumber \\
&+ 4 h a_1 \left|a_3\right|^2 d_\mathrm{eff} n_3 \omega_2 \left(1 + i \Delta k \, x - e^{i \Delta k \, x}\right) \Big\} \nonumber \, \\
e_{22} &= \frac{h d_\mathrm{eff} \omega_2}{c^2 \left(\Delta k\right)^2 n_1 n_2 n_3} \Big\{n_1 a_{1}^\ast \Big[2 a_3 n_3 c \Delta k \left(1 + h\right) \left(1 - e^{i \Delta k \, x}\right) \\
&+ 4 h a_1 a_2 d_\mathrm{eff} \omega_3 \left(e^{i \Delta k \, x} - i \Delta k\, x - 1\right)\Big] \nonumber \\
&+ 4 h a_2 \left|a_3\right|^2 d_\mathrm{eff} n_3 \omega_1 \left(1 + i \Delta k \, x - e^{i \Delta k \, x}\right) \Big\} \nonumber \\
e_{32} &= \frac{h d_\mathrm{eff} \omega_3}{c^2 \left(\Delta k\right)^2 n_1 n_2 n_3} e^{-i \Delta k \, x} \Big\{2 a_1 a_2 n_1 n_2 c \Delta k (1 + h) \left(1 - e^{i \Delta k x}\right) \\
&+ 4 a_3 d_\mathrm{eff} h \left[1 + e^{i \Delta k\, x} \left(i \Delta k\, x - 1\right)\right] \left(\left|a_1\right|^2 n_1 \omega_2 +
\left|a_2\right|^2 n_2 \omega_1\right) \Big\} \nonumber
\end{align}
The third-order deformation are given by
\begin{align}
e_{13} &= \frac{d_\mathrm{eff} h \omega_1}{c^3 \left(\Delta k\right)^3 n_{1}^2 n_2 n_3} \Bigg\{ 8 a_2 a_3 d_\mathrm{eff}^2 h^2 n_2 \omega_1 \omega_3 \left[e^{j \Delta k\, x} \left(2 - j Dk x\right) -2 - j\Delta k \,x\right] \left(a_{2}^\ast\right)^2 \\
&+ a_{2}^\ast \Big[2 c \Delta k\, \left(1 + h\right) n_1 n_2 \Big(a_3 c \Delta k \,\left(1 - e^{j \Delta k\, x}\right) \left(1 + h\right) n_3 +  4 a_1 a_2 d_\mathrm{eff} h \omega_3 \nonumber \\
&\times \left(e^{j \Delta k\, x} - j \Delta k\, x - 1\right)\Big) \nonumber \\
&+ 8 a_3 d_\mathrm{eff}^2 h^2 \omega_2 \left(2 + j \Delta k\, x + e^{j \Delta k\, x} (j \Delta k\, x - 2)\right) \left(a_3 n_3 \omega_1 a_{3}^\ast - 2 a_1 n_1 \omega_3 a_{1}^\ast \right) \nonumber \\
&+8 a_1 d_\mathrm{eff} h n_1 \omega_2 a_{3}^\ast \Big(a_3 c \Delta k \, (1 + h) n_3 \left(1 + j \Delta k\, x - e^{j \Delta k\, x}\right) \nonumber \\
&-2 j a_1 a_2 d_\mathrm{eff} \Delta k \, h \omega_3 x + 2 j a_1 a_2 d_\mathrm{eff} h \omega_3 \sin\left(\Delta k\, x\right)\Big) \Big] \Bigg\} \, , \nonumber \\
e_{23} &= \frac{d_\mathrm{eff} h \omega_2}{c^3 \left(\Delta k\right)^3 n_{1} n_{2}^2 n_3} \Bigg\{ 8 a_1 a_3 d_\mathrm{eff}^2 h^2 n_1 \omega_2 \omega_3 \left[e^{j \Delta k\, x} \left(2 - j Dk x\right) -2 - j\Delta k \,x\right] \left(a_{1}^\ast\right)^2 \\
&+ a_{1}^\ast \Big[2 c \Delta k\, \left(1 + h\right) n_1 n_2 \Big(a_3 c \Delta k \,\left(1 - e^{j \Delta k\, x}\right) \left(1 + h\right) n_3 +  4 a_1 a_2 d_\mathrm{eff} h \omega_3 \nonumber \\
&\times \left(e^{j \Delta k\, x} - j \Delta k\, x - 1\right)\Big) \nonumber \\
&+ 8 a_3 d_\mathrm{eff}^2 h^2 \omega_1 \left(2 + j \Delta k\, x + e^{j \Delta k\, x} (j \Delta k\, x - 2)\right) \left(a_3 n_3 \omega_2 a_{3}^\ast - 2 a_2 n_2 \omega_3 a_{2}^\ast \right) \nonumber \\
&+8 a_2 d_\mathrm{eff} h n_2 \omega_1 a_{3}^\ast \Big(a_3 c \Delta k \, (1 + h) n_3 \left(1 + j \Delta k\, x - e^{j \Delta k\, x}\right) \nonumber \\
&-2 j a_1 a_2 d_\mathrm{eff} \Delta k \, h \omega_3 x + 2 j a_1 a_2 d_\mathrm{eff} h \omega_3 \sin\left(\Delta k\, x\right)\Big) \Big] \Bigg\} \, , \nonumber \\
e_{33} &= \frac{d_\mathrm{eff} h \omega_3}{c^3 \left(\Delta k\right)^3 n_{1} n_{2} n_{3}^2} e^{-j \Delta k \, x} \Bigg\{ -a_2 \Big[2 a_1 c^2 \left(\Delta k\right)^2 \left(e^{j \Delta k \, x} - 1\right) \left(1 + h\right)^2 n_1 n_2 n_3 \\
&+ 2 d_\mathrm{eff} h \Big(4 a_{2}^\ast n_2 \omega_1 \Big(a_3 c \Delta k \, \left(1 + h\right) n_3 \left(e^{j \Delta k \, x} \left(1 - j \Delta k \, x\right) - 1\right) \nonumber \\
&+ a_1 a_2 d_\mathrm{eff} h \omega_3 \left(2 + j \Delta k \, x + e^{j \Delta k \, x} \left(j \Delta k \, x - 2\right)\right)\Big) \nonumber \\
&+ 4 a_1 d_\mathrm{eff} h \omega_2 \left(2 + j \Delta k \, x + e^{j \Delta k \, x} \left(j \Delta k \, x - 2\right)\right) \left(a_1 n_1 \omega_3 a_{1}^\ast - 2 a_3 n_3 \omega_1 a_{3}^\ast\right)\Big)\Big] \nonumber \\
&+8 a_3 d_\mathrm{eff} h n_3 \omega_2 a_{1}^\ast \Big[a_1 c \Delta k \, \left(1 + h\right) n_1 \left(1 + e^{j \Delta k \, x} \left(j \Delta k \, x - 1\right)\right) \nonumber \\
&+ 2 j a_3 d_\mathrm{eff} e^{j \Delta k \, x} h \omega_1 a_{2}^\ast \left(\Delta k \, x - \sin\left(\Delta k \, x\right)\right) \Big] \Bigg\} \, . \nonumber
\end{align}

\section{Perfectly phase matched HAM terms}
\label{app:match}

The HAM terms for the special case of perfect phase matching for three-wave mixing follows, where the expressions are generated after letting $\Delta k \rightarrow 0$. The initial guesses are the same as in Eqs. \ref{eq:e10calc}-\ref{eq:e30calc}. The first-order terms are given by
\begin{align}
e_{11} &= -\frac{2j}{c n_1} a_{2}^\ast a_3 d_\mathrm{eff} h \omega_1 x \, , \\
e_{21} &= -\frac{2j}{c n_2} a_{1}^\ast a_3 d_\mathrm{eff} h \omega_2 x \, , \\
e_{31} &= -\frac{2j}{c n_3} a_1 a_{2} d_\mathrm{eff} h \omega_3 x \, .
\end{align}
The second-order terms are given by
\begin{align}
e_{12} &= \frac{2 d_\mathrm{eff} h \omega_1 x}{c^2 n_1 n_2 n_3} \Big[ a_1 \left|a_3\right|^2 d_\mathrm{eff} h n_3 \omega_2 x - n_2 a_{2}^\ast \left(j a_3 c \left(1 + h\right) n_3 + a_1 a_2 d_\mathrm{eff} h \omega_3 x\right) \Big] \, , \\
e_{22} &= \frac{2 d_\mathrm{eff} h \omega_2 x}{c^2 n_1 n_2 n_3} \Big[ a_2 \left|a_3\right|^2 d_\mathrm{eff} h n_3 \omega_1 x - n_1 a_{1}^\ast \left(j a_3 c \left(1 + h\right) n_3 + a_1 a_2 d_\mathrm{eff} h \omega_3 x\right) \Big] \, , \\
e_{32} &= -\frac{2 d_\mathrm{eff} h \omega_3 x}{c^2 n_1 n_2 n_3} \Big[ j a_1 a_2 c (1 + h) n_1 n_2 + a_3 d_\mathrm{eff} h x \left(\left|a_1\right|^2 n_1 \omega_2 + \left|a_2\right|^2 n_2 \omega_1\right) \Big] \, .
\end{align}
The third-order terms are given by
\begin{align}
e_{13} &= \frac{2 j d_\mathrm{eff} h \omega_1 x}{3 c^3 n_{1}^2 n_2 n_3} \Big\{ 2 \left|a_2\right|^2 a_{2}^\ast a_3 d_{\mathrm{eff}}^2 h^2 n_2 \omega_1 \omega_3 x^2 \\
&- 2 a_1 a_{3}^\ast d_\mathrm{eff} h n_1 \omega_2 x \left(3 j a_3 c \left(1 + h\right) n_3 + 2 a_1 a_2 d_\mathrm{eff} h \omega_3 x\right) \nonumber \\
&- a_{2}^\ast \Big[3 c \left(1 + h\right) n_1 n_2 \left(a_3 c \left(1 + h\right) n_3 - 2 j a_1 a_2 d_\mathrm{eff} h \omega_3 x\right) \nonumber \\
&+ 2 a_3 d_{\mathrm{eff}}^2 h^2 \omega_2 x^2 \left(\left|a_3\right|^2 n_3 \omega_1 - 2 \left|a_1\right|^2 n_1 \omega_3\right) \Big] \Big\} \nonumber \, , \\
e_{23} &= \frac{2 j d_\mathrm{eff} h \omega_2 x}{3 c^3 n_{1} n_{2}^2 n_3} \Big\{ 2 \left|a_1\right|^2 a_{1}^\ast a_3 d_{\mathrm{eff}}^2 h^2 n_1 \omega_2 \omega_3 x^2 \\
&- 2 a_2 a_{3}^\ast d_\mathrm{eff} h n_2 \omega_1 x \left(3 j a_3 c \left(1 + h\right) n_3 + 2 a_1 a_2 d_\mathrm{eff} h \omega_3 x\right) \nonumber \\
&- a_{1}^\ast \Big[3 c \left(1 + h\right) n_1 n_2 \left(a_3 c \left(1 + h\right) n_3 - 2 j a_1 a_2 d_\mathrm{eff} h \omega_3 x\right) \nonumber \\
&+ 2 a_3 d_{\mathrm{eff}}^2 h^2 \omega_1 x^2 \left(\left|a_3\right|^2 n_3 \omega_2 - 2 \left|a_2\right|^2 n_2 \omega_3\right) \Big] \Big\} \nonumber \, , \\
e_{33} &= \frac{2 j d_\mathrm{eff} h \omega_3 x}{3 c^3 n_{1} n_2 n_{3}^2} \Big\{ 2 d_\mathrm{eff} h x \Big[a_1 a_2 d_\mathrm{eff} h n_1 \omega_2 \omega_3 x \left|a1\right|^2 \\
&+  a_1 a_2 d_\mathrm{eff} h n_2 \omega_1 \omega_3 x \left|a2\right|^2 + a_3 n_3 \Big(\omega_2 a_{1}^\ast \left(3 j a_1 c (1 + h) n_1 + 2 a_3 d_\mathrm{eff} h \omega_1 x a_{2}^\ast\right) \nonumber \\
&+ a_2 \omega_1 \left(3 j c (1 + h) n_2 a_{2}^\ast - 2 a_1 d_\mathrm{eff} h \omega_2 x a_{3}^\ast\right)\Big) \Big] - 3 a_1 a_2 c^2 (1 + h)^2 n_1 n_2 n_3 \Big\} \nonumber
\end{align}
The fourth-order terms are given by
\begin{align}
e_{14} &= \frac{d_\mathrm{eff} h \omega_1 x}{96 c^4 n_{1}^2 n_{2}^2 n_{3}^2} \Big\{ 64 d_{\mathrm{eff}}^2 h^2 n_2 \omega_1 \omega_3 x^2 \Big[a_1 \left|a_2\right|^4 d_\mathrm{eff} h n_2 \omega_3 x \\
&+ 2 \left(a_{2}^\ast\right)^2 a_3 n_3 \left(3 j a_2 c \left(1 + h\right) n_2 + 2 a_{1}^\ast a_3 d_\mathrm{eff} h \omega_2 x\right)\Big] \nonumber \\
&+ 4 a_1 a_{3}^\ast d_\mathrm{eff} h n_3 \omega_2 x \Big[48 c \left(1 + h\right) n_1 n_2 \left(3 a_3 c (1 + h) n_3 - 4 j a_1 a_2 d_\mathrm{eff} h \omega_3 x\right) \nonumber \\
&+ 16 a_3 d_{\mathrm{eff}}^2 h^2 \omega_2 x^2 \left(\left|a_3\right|^2 n_3 \omega_1 - 4 \left|a_1\right|^2 n_1 \omega_3\right) \Big] \nonumber \\
&- 2 j a_{2}^\ast n_2 \Big[96 c^2 (1 + h)^2 n_1 n_2 n_3 \left(a_3 c \left(1 + h\right) n_3 - 3 j a_1 a_2 d_\mathrm{eff} h \omega_3 x\right) \nonumber \\
&+ 64 j d_{\mathrm{eff}}^2 h^2 \omega_2 x^2 \Big(2 a_1 \left|a1\right|^2 a_2 d_\mathrm{eff} h n_1 \omega_{3}^2 x \nonumber \\
&- a_3 n_3 \big(5 a_1 a_2 a_{3}^\ast d_\mathrm{eff} h \omega_1 \omega_3 x + 3 j c \left(1 + h\right) \big(\left|a3\right|^2 n_3 \omega_1 - 2 \left|a_1\right|^2 n_1 \omega_3\big)\big) \Big) \Big] \Big\} \, , \nonumber \\
e_{24} &= \frac{d_\mathrm{eff} h \omega_2 x}{96 c^4 n_{1}^2 n_{2}^2 n_{3}^2} \Big\{ 64 d_{\mathrm{eff}}^2 h^2 n_1 \omega_2 \omega_3 x^2 \Big[ \left|a_1\right|^4 a_2 d_\mathrm{eff} h n_1 \omega_3 x \\
&+ 2 \left(a_{2}^\ast\right)^2 a_3 n_3 \left(3 j a_1 c \left(1 + h\right) n_1 + 2 a_{2}^\ast a_3 d_\mathrm{eff} h \omega_1 x \right)\Big] \nonumber \\
&+ 4 a_2 a_{3}^\ast d_\mathrm{eff} h n_3 \omega_1 x \Big[48 c \left(1 + h\right) n_1 n_2 \left(3 a_3 c (1 + h) n_3 - 4 j a_1 a_2 d_\mathrm{eff} h \omega_3 x\right) \nonumber \\
&+ 16 a_3 d_{\mathrm{eff}}^2 h^2 \omega_1 x^2 \left(\left|a_3\right|^2 n_3 \omega_2 - 4 \left|a_2\right|^2 n_2 \omega_3\right) \Big] \nonumber \\
&- 2 j a_{1}^\ast n_1 \Big[96 c^2 (1 + h)^2 n_1 n_2 n_3 \left(a_3 c \left(1 + h\right) n_3 - 3 j a_1 a_2 d_\mathrm{eff} h \omega_3 x\right) \nonumber \\
&+ 64 j d_{\mathrm{eff}}^2 h^2 \omega_1 x^2 \Big(2 a_1 a_2 \left|a2\right|^2 d_\mathrm{eff} h n_2 \omega_{3}^2 x \nonumber \\
&- a_3 n_3 \big(5 a_1 a_2 a_{3}^\ast d_\mathrm{eff} h \omega_2 \omega_3 x + 3 j c \left(1 + h\right) \big(\left|a3\right|^2 n_3 \omega_2 - 2 \left|a_2\right|^2 n_2 \omega_3\big)\big) \Big) \Big] \Big\} \, , \nonumber \\
e_{34} &= - \frac{2 d_\mathrm{eff} h \omega_3 x}{3 c^4 n_{1}^2 n_{2}^2 n_{3}^2} \Big\{ 3 j a_1 a_2 c^3 \left(1 + h\right)^3 n_{1}^2 n_{2}^2 n_3 \\
&- d_\mathrm{eff} h x \Big[6 j a_1 \left|a_1\right|^2 a_2 c d_\mathrm{eff} h \left(1 + h\right) n_{1}^2 n_2 \omega_2 \omega_3 x + \left|a_1\right|^4 a_3 d_{\mathrm{eff}}^2 h^2 n_{1}^2 \omega_{2}^2 \omega_3 x^2 \nonumber \\
&+ 6 j a_1 a_2 \left|a_2\right|^2 c d_\mathrm{eff} h \left(1 + h\right) n_1 n_{2}^2 \omega_1 \omega_3 x + \left|a_2\right|^4 a_3 d_{\mathrm{eff}}^2 h^2 n_{2}^2 \omega_{1}^2 \omega_3 x^2 \nonumber \\
&- a_{1}^\ast a_3 n_1 \omega_2 \Big(9 a_1 c^2 \left(1 + h\right)^2 n_1 n_2 n_3 - 2 d_\mathrm{eff} h \omega_1 x \big(6 j a_{2}^\ast a_3 c \left(1 + h\right) n_2 n_3 \nonumber \\
&+ 5 a_1 \left|a_2\right|^2 d_\mathrm{eff} h n_2 \omega_3 x - 2 a_1 \left|a_3\right|^2 d_\mathrm{eff} h n_3 \omega_2 x\big) \Big) \nonumber \\
&-a_2 n_2 \omega_1 \Big(4 a_1 a_{3}^\ast d_\mathrm{eff} h n_1 \omega_2 x \left(3 j a_3 c \left(1 + h\right) n_3 + a_1 a_2 d_\mathrm{eff} h \omega_3 x\right) \nonumber \\
&+ 9 a_{2}^\ast a_3 c^2 (1 + h)^2 n_1 n_2 n_3 + 4 a_{2}^\ast a_3 \left|a_3\right|^2 d_{\mathrm{eff}}^2 h^2 n_3 \omega_1 \omega_2 x^2 \Big) \Big] \Big\} \nonumber \, .
\end{align}

\section*{Acknowledgments}
NJD thanks the Hawaii Pacific University, College of Natural and Computational Sciences, Scholarly Endeavors Program for their continued support. MK thanks the University of The Bahamas Internal Grants Programme for Research, Creative and Artistic Proposals for their support.



\end{document}